\pdfoutput=1
\documentclass[12pt,a4paper]{article}
\usepackage{ifthen}
\usepackage{booktabs}
\usepackage{overpic}
\usepackage{relsize}
\usepackage{rotating}
\usepackage[top=1in, bottom=1.25in, left=1in, right=1in]{geometry}

\newboolean{pdflatex}
\setboolean{pdflatex}{true}

\newboolean{articletitles}
\setboolean{articletitles}{true}

\newboolean{uprightparticles}
\setboolean{uprightparticles}{false}

\newboolean{inbibliography}
\setboolean{inbibliography}{false}

\def\paperauthors{LHCb collaboration}
\def\paperasciititle{Search for dark photons produced in 13 TeV pp collisions}
\def\papertitle{Search for dark photons produced in 13\tev $pp$ collisions}
\def\paperkeywords{{High Energy Physics}, {LHCb}}
\def\papercopyright{CERN on behalf of the LHCb collaboration}
\def\paperlicence{CC-BY-4.0}
\def\paperlicenceurl{https://creativecommons.org/licenses/by/4.0/}

\usepackage{microtype}
\usepackage{lineno}
\usepackage{xspace}
\usepackage{caption}

\usepackage{graphicx}
\usepackage{color}
\usepackage{colortbl}
\graphicspath{{./figs/}}

\usepackage{amsmath}
\usepackage{amssymb}
\usepackage{amsfonts}
\usepackage{upgreek}

\newcommand*\patchAmsMathEnvironmentForLineno[1]{%
\expandafter\let\csname old#1\expandafter\endcsname\csname #1\endcsname
\expandafter\let\csname oldend#1\expandafter\endcsname\csname
end#1\endcsname
 \renewenvironment{#1}%
   {\linenomath\csname old#1\endcsname}%
   {\csname oldend#1\endcsname\endlinenomath}%
}
\newcommand*\patchBothAmsMathEnvironmentsForLineno[1]{%
  \patchAmsMathEnvironmentForLineno{#1}%
  \patchAmsMathEnvironmentForLineno{#1*}%
}
\AtBeginDocument{%
\patchBothAmsMathEnvironmentsForLineno{equation}%
\patchBothAmsMathEnvironmentsForLineno{align}%
\patchBothAmsMathEnvironmentsForLineno{flalign}%
\patchBothAmsMathEnvironmentsForLineno{alignat}%
\patchBothAmsMathEnvironmentsForLineno{gather}%
\patchBothAmsMathEnvironmentsForLineno{multline}%
\patchBothAmsMathEnvironmentsForLineno{eqnarray}%
}

\usepackage{hyperxmp}

\usepackage[pdftex,
            pdfauthor={\paperauthors},
            pdftitle={\paperasciititle},
            pdfkeywords={\paperkeywords},
            pdfcopyright={Copyright (C) \papercopyright},
            pdflicenseurl={\paperlicenceurl}]{hyperref}

\usepackage[all]{hypcap} 

\usepackage{cite}
\usepackage{mciteplus}

\def\lhcb {\mbox{LHCb}\xspace}

\newcommand{\tev}{\ifthenelse{\boolean{inbibliography}}{\ensuremath{~T\kern -0.05em eV}\xspace}{\ensuremath{\mathrm{\,Te\kern -0.1em V}}}\xspace}
\newcommand{\gev}{\ensuremath{\mathrm{\,Ge\kern -0.1em V}}\xspace}
\newcommand{\mev}{\ensuremath{\mathrm{\,Me\kern -0.1em V}}\xspace}
\def\invfb   {\ensuremath{\mbox{\,fb}^{-1}}\xspace}
\def \xip {\ensuremath{\chi^2_{\mathrm{IP}}(\mu)}\xspace}
\def \mxip {\ensuremath{{\rm min}[\chi^2_{\mathrm{IP}}(\mu^{\pm})]}\xspace}
\def \xv {\ensuremath{\chi^2_{\mathrm{VF}}(\mu^+\mu^-)}\xspace}
\def\aprime{\ensuremath{A^{\prime}}\xspace}
\def\atomm{\ensuremath{\aprime\!\to\!\mu^+\mu^-}\xspace}
\def\gtomm{\ensuremath{\gamma^*\!\to\!\mu^+\mu^-}\xspace}
\def \mmm {\ensuremath{m(\mu^+\mu^-)}\xspace}
\def\ps   {\ensuremath{\mbox{\,ps}}\xspace}
\def\ma{\ensuremath{m(\aprime)}\xspace}
\def\ta{\ensuremath{\tau(\aprime)}\xspace}
\def\pt         {\mbox{$p_{\rm T}$}\xspace}
\def\KS      {{\ensuremath{K^0_{\mathrm{ \scriptscriptstyle S}}}}\xspace}

\newcommand{\chisq}{\ensuremath{\chi^2}\xspace}
\def\sa{\ensuremath{\sigma[\mmm]}\xspace}
\def\jpsi     {{\ensuremath{{J\mskip -3mu/\mskip -2mu\psi\mskip 2mu}}}\xspace}
\mathchardef\Upsilon="7107

\mathchardef\PLambda="7103
\def\ngob {\ensuremath{n_{\rm ob}^{\gamma^*}[\ma]}\xspace}
\def\naob {\ensuremath{n_{\rm ob}^{\aprime}[\ma]}\xspace}
\def\naex {\ensuremath{n_{\rm ex}^{\aprime}[\ma,\varepsilon^2]}\xspace}
\def\naobt {\ensuremath{n_{\rm ob}^{\aprime}[\ma,\tau(\aprime)]}\xspace}
\def\naobe {\ensuremath{n_{\rm ob}^{\aprime}[\ma,\varepsilon^2]}\xspace}
\def\effr {\ensuremath{\epsilon_{{}^{\gamma^*}}^{{}_{\aprime}}[\ma,\ta]}\xspace}
\def\maeps {\ensuremath{[\ma,\varepsilon^2]}\xspace}

\begin{document}

\renewcommand{\thefootnote}{\fnsymbol{footnote}}
\setcounter{footnote}{1}

\begin{titlepage}
\pagenumbering{roman}

\vspace*{-1.5cm}
\centerline{\large EUROPEAN ORGANIZATION FOR NUCLEAR RESEARCH (CERN)}
\vspace*{1.5cm}
\noindent
\begin{tabular*}{\linewidth}{lc@{\extracolsep{\fill}}r@{\extracolsep{0pt}}}
\vspace*{-2.7cm}\mbox{\!\!\!\includegraphics[width=.14\textwidth]{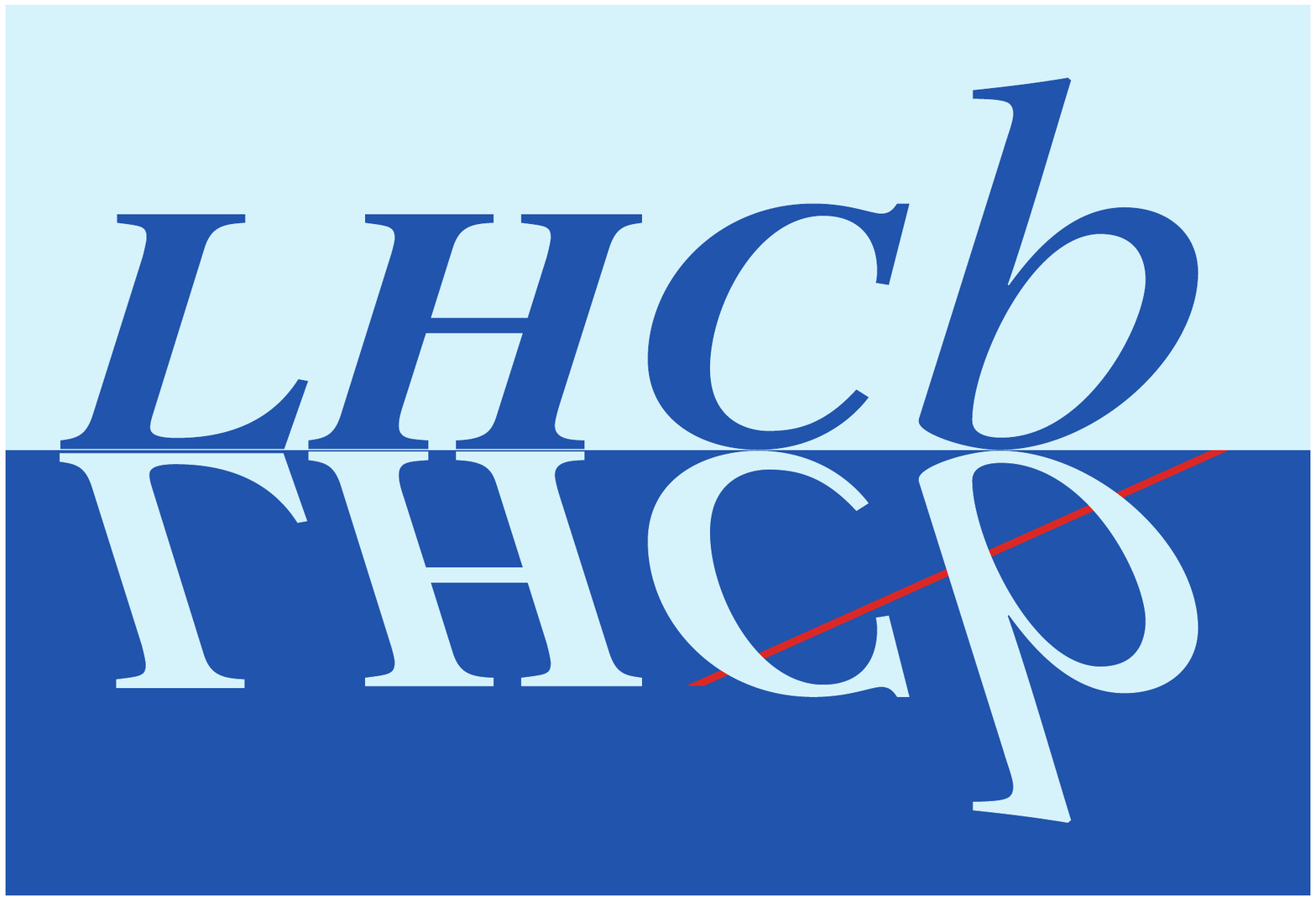}} & &
\\
 & & CERN-EP-2017-248 \\  
 & & LHCb-PAPER-2017-038 \\  
 & & October 5, 2017 \\
 & & \\
\end{tabular*}

\vspace*{4.0cm}

{\normalfont\bfseries\boldmath\huge
\begin{center}
  \papertitle
\end{center}
}

\vspace*{2.0cm}

\begin{center}
\paperauthors\footnote{Authors are listed at the end of this Letter.}
\end{center}

\vspace{\fill}

\begin{abstract}
  \noindent
Searches are performed for both prompt-like and long-lived dark photons, \aprime, produced in proton-proton collisions at a center-of-mass energy of 13\tev, using \atomm decays and a data sample corresponding to an integrated luminosity of 1.6\invfb collected with the LHCb detector.
The prompt-like \aprime search covers the mass range from near the dimuon threshold up to 70\gev, while the long-lived \aprime search is restricted to the low-mass region $214<\ma<350\mev$.
No evidence for a signal is found, and 90\% confidence level exclusion limits are placed on the $\gamma$--$\aprime$ kinetic-mixing strength.
The constraints placed on prompt-like dark photons are the most stringent to date for the mass range $10.6 < \ma < 70\gev$, and are comparable to the best existing limits for $\ma < 0.5\gev$.
The search for long-lived dark photons is the first to achieve sensitivity using a displaced-vertex signature.
\end{abstract}

\vspace*{2.0cm}

\begin{center}
  Published in Physical Review Letters {\bf 120}, 061801 (2018)
\end{center}

\vspace{\fill}

{\footnotesize
\centerline{\copyright~\papercopyright, licence \href{\paperlicenceurl}{\paperlicence}.}}
\vspace*{2mm}

\end{titlepage}

\newpage
\setcounter{page}{2}
\mbox{~}

\cleardoublepage

\renewcommand{\thefootnote}{\arabic{footnote}}
\setcounter{footnote}{0}

\pagestyle{plain}
\setcounter{page}{1}
\pagenumbering{arabic}

The possibility that dark matter particles may interact via unknown forces, felt only feebly by Standard Model (SM) particles, has motivated substantial effort to search for dark-sector forces (see Ref.~\cite{Alexander:2016aln} for a review).
A compelling dark-force scenario involves a massive {\em dark photon}, \aprime, whose coupling to the electromagnetic current
is suppressed relative to that of the ordinary photon, $\gamma$, by a factor of $\varepsilon$.
In the minimal model, the dark photon does not couple directly to charged SM particles; however, a coupling may arise via kinetic mixing between the SM hypercharge and \aprime field strength tensors~\cite{Okun:1982xi,Galison:1983pa,Holdom:1985ag,Pospelov:2007mp,ArkaniHamed:2008qn,Bjorken:2009mm}.
This mixing provides a potential portal through which dark photons may be produced if kinematically allowed.
If the kinetic mixing arises due to processes whose amplitudes involve one or two loops containing high-mass particles, perhaps even at the Planck scale, then $10^{-12} \lesssim \varepsilon^2 \lesssim 10^{-4}$ is expected~\cite{Alexander:2016aln}.
Fully exploring this {\em few-loop} range of kinetic-mixing strength is an important goal of dark-sector physics.

Constraints have been placed  on visible \aprime decays by previous beam-dump~\cite{Bergsma:1985is,Konaka:1986cb,Riordan:1987aw,Bjorken:1988as,Bross:1989mp,Davier:1989wz,Athanassopoulos:1997er,Astier:2001ck,Adler:2004hp,Bjorken:2009mm,Artamonov:2009sz,Essig:2010gu,Blumlein:2011mv,Gninenko:2012eq,Blumlein:2013cua},
fixed-target~\cite{Abrahamyan:2011gv,Merkel:2014avp,Merkel:2011ze},
collider~\cite{Aubert:2009cp,Curtin:2013fra,Lees:2014xha,Ablikim:2017aab},
and rare-meson-decay \cite{Bernardi:1985ny,MeijerDrees:1992kd,Archilli:2011zc,Gninenko:2011uv,Babusci:2012cr,Adlarson:2013eza,Agakishiev:2013fwl,Adare:2014mgk,Batley:2015lha,KLOE:2016lwm}
experiments.
The few-loop region is ruled out for dark photon masses $\ma \lesssim 10\mev$ ($c=1$ throughout this Letter).
Additionally, the region $\varepsilon^2 \gtrsim 5\!\times\!10^{-7}$ is excluded for $\ma < 10.2\gev$, along with about half of the remaining few-loop region below the dimuon threshold.
Many ideas have been proposed to further explore the
\maeps parameter space~\cite{Essig:2010xa,
Freytsis:2009bh,Balewski:2013oza,
Wojtsekhowski:2012zq,
Beranek:2013yqa,
Echenard:2014lma,
Battaglieri:2014hga,
Alekhin:2015byh,
Gardner:2015wea,
Ilten:2015hya,
Curtin:2014cca,
He:2017ord,Kozaczuk:2017per},
including an inclusive search for \atomm decays with the LHCb experiment, which is predicted to provide sensitivity to large regions of otherwise inaccessible parameter space using data to be collected during Run~3 of the LHC (2021--2023)~\cite{Ilten:2016tkc}.

A dark photon produced in proton-proton, $pp$, collisions via $\gamma$--\aprime mixing inherits the production mechanisms of an off-shell photon with $m(\gamma^*) = \ma$;
therefore, both the production and decay kinematics of the \atomm and \gtomm processes are identical.
Furthermore, the expected \atomm signal yield  is given by~\cite{Ilten:2016tkc}
\begin{equation}
  \label{eq:norm}
\naex = \varepsilon^2 \left[\frac{\ngob}{2\Delta m}\right] \mathcal{F}[\ma]\, \effr,
\end{equation}
where \ngob is the observed prompt \gtomm yield
 in a small $\pm\Delta m$ window around \ma,
the function $\mathcal{F}[\ma]$ includes phase-space and other known factors,
and \effr is the ratio of the \atomm and \gtomm detection efficiencies, which depends on the \aprime lifetime, \ta.
If \aprime decays to invisible final states are negligible, then $\ta \propto [\ma \varepsilon^2]^{-1}$ and \atomm decays can potentially be reconstructed as displaced from the primary $pp$ vertex (PV) when the product $\ma\varepsilon^2$ is small.
When \ta is small compared to the experimental resolution, \atomm decays are reconstructed as prompt-like and are experimentally indistinguishable from prompt \gtomm production, resulting in $\effr \approx 1$.
This facilitates a fully data-driven search and the cancelation of most experimental systematic effects, since the observed \atomm yields, \naob, can be normalized to \naex to obtain constraints on $\varepsilon^2$.

This Letter presents searches for both prompt-like and long-lived dark photons produced in $pp$ collisions at a center-of-mass energy of 13\tev, using \atomm decays and a data sample corresponding to an integrated luminosity of 1.6\invfb  collected with the LHCb detector in 2016.
The prompt-like \aprime search is performed from near the dimuon threshold up to 70\gev, above which the \mmm spectrum is dominated by the $Z$ boson.
The long-lived \aprime search
is restricted to the mass range $214<\ma<350\mev$, where the data sample potentially provides sensitivity.

The \lhcb detector is a single-arm forward spectrometer covering the pseudorapidity range $2<\eta <5$, described in detail in Refs.~\cite{Alves:2008zz,LHCb-DP-2014-002}.
Simulated data samples, which are used to validate the analysis, are produced using the software described in Refs.~\cite{Sjostrand:2014zea,*Sjostrand:2007gs,LHCb-PROC-2010-056,Allison:2006ve, *Agostinelli:2002hh}.
The online event selection is performed by a trigger~\cite{LHCb-DP-2012-004}, which consists of a hardware stage using information from the calorimeter and muon systems, followed by a software stage, which performs a full event
reconstruction.
At the hardware stage, events are required to have a muon with $\pt \gtrsim 1.8\gev$, where \pt is the momentum transverse to the beam direction,
or a dimuon in which the product of the \pt of each muon is in excess of $(\approx\!1.5\gev)^2$.
The long-lived \aprime search also uses events selected at the hardware stage independently of the \atomm candidate.
In the software stage, \atomm candidates are built from two oppositely charged tracks
that form a good quality vertex and satisfy stringent muon-identification criteria.
The muons are required to have $2<\eta<4.5$,
$\pt > 0.5\,(1.0)\gev$, momentum ${p > 10\,(20)\gev}$,
and be inconsistent (consistent) with originating from the PV in the long-lived (prompt-like) \aprime search.
Finally, the \aprime candidates are required to
satisfy $\pt > 1\gev$, $2 < \eta<4.5$, and
have a decay topology consistent with originating from the PV.

The prompt-like \aprime search is based on a data sample where all online-reconstructed particles are stored, but most lower-level information is discarded, greatly reducing the event size.
This data-storage strategy, made possible by advances in the LHCb data-taking scheme introduced in 2015~\cite{LHCb-PROC-2015-011,Aaij:2016rxn},
 permits the recording of all events that contain a prompt-like dimuon candidate without placing any requirements on \mmm.
 The \mmm spectrum recorded by the trigger is provided in the Supplemental Material to this Letter~\cite{Supp}.

Three main types of background contribute to the prompt-like \aprime search:
prompt ${\gtomm}$ production, which is irreducible;
resonant decays to $\mu^+\mu^-$, whose mass-peak regions are avoided in the search;
and
various types of misreconstruction.
The misreconstruction background consists of three dominant contributions:
double misidentification of prompt hadrons as muons, $hh$;
a misidentified prompt hadron combined with a muon produced in a decay of a hadron containing a heavy-flavor quark, $Q$, where the muon is misreconstructed as prompt-like, $h\mu_Q$;
and the misreconstruction of two muons produced in $Q$-hadron decays, $\mu_Q\mu_Q$.
These backgrounds are highly suppressed by the stringent muon-identification and prompt-like requirements applied in the trigger; however, in the region $[m(\phi),m(\Upsilon)]$, the misreconstructed backgrounds overwhelm the signal-like \gtomm contribution.

\begin{figure}
  \centering
  \includegraphics[width=0.99\textwidth]{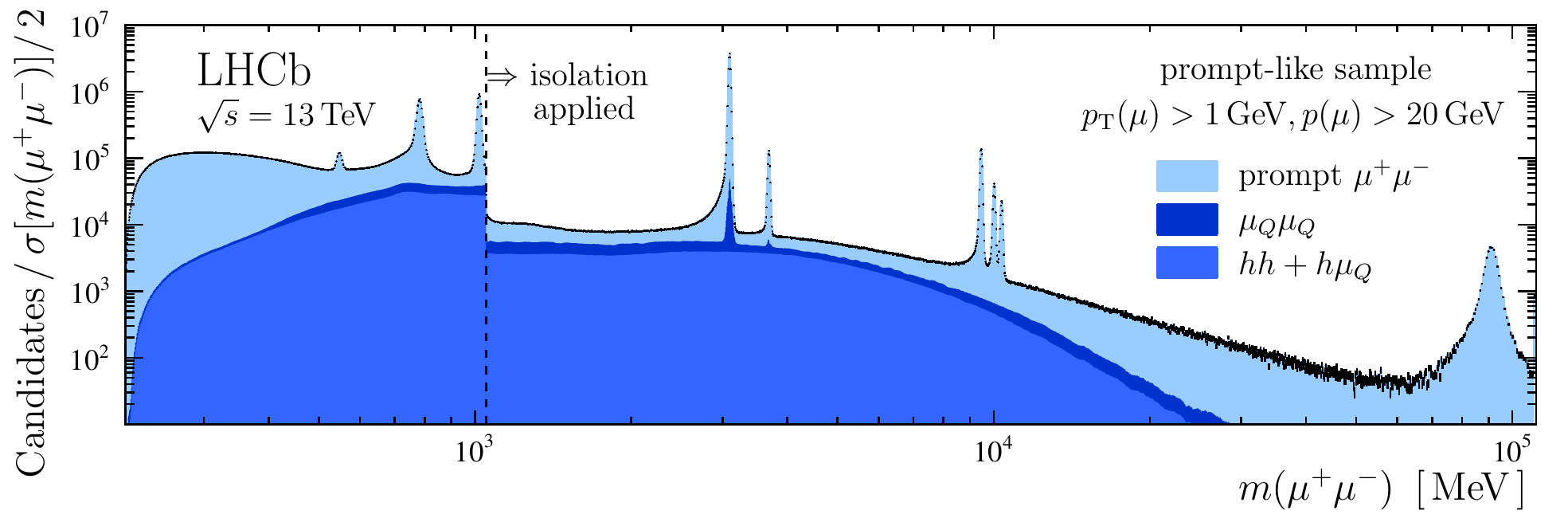}
  \caption{
  Prompt-like mass spectrum,
  where the categorization of the data as prompt $\mu^+\mu^-$, $\mu_Q\mu_Q$, and $hh+h\mu_Q$ is determined using the fits described in the text.
  }
  \label{fig:m_prompt}
\end{figure}

For masses below (above) the $\phi$ meson mass, dark photons are expected to be predominantly produced in meson-decay (Drell-Yan) processes in $pp$ collisions at LHCb.
A well-known signature of Drell-Yan production is dimuons that are largely isolated, and a high-mass dark photon would inherit this property.
The signal sensitivity is enhanced by applying a jet-based isolation requirement for $\ma > m(\phi)$, which improves the sensitivity by up to a factor of two at low masses and by $\mathcal{O}(10\%)$ for $\ma > 10\gev$.
Jet reconstruction is performed
 by clustering charged and neutral particle-flow candidates~\cite{LHCb-PAPER-2013-058}
using the anti-$k_{\rm T}$ clustering algorithm~\cite{antikt} with $R=0.5$ as implemented in \textsc{FastJet}~\cite{fastjet}.
Muons with $\pt(\mu)/\pt({\rm jet}) < 0.7$ are rejected, where the contribution to $\pt({\rm jet})$ from the other muon is excluded if both muons are clustered in the same jet,
as this is found to provide nearly optimal sensitivity for all $\ma > m(\phi)$.
Figure~\ref{fig:m_prompt} shows the resulting prompt-like \mmm spectrum using $\Delta m$ bins that are \sa/2 wide, where \sa is the mass resolution which varies from about 0.7\mev near threshold to 0.7\gev at $\mmm=70\gev$.

The prompt-like \aprime search strategy involves determining the observed \atomm yields from fits to the \mmm spectrum,
and normalizing them using Eq.~\ref{eq:norm} to obtain constraints on $\varepsilon^2$.
To determine \ngob for use in Eq.~\ref{eq:norm}, binned extended maximum likelihood fits are performed using the dimuon vertex-fit quality, \xv,
and \mxip distributions, where \xip is defined as the difference in $\chi^2_{\rm VF}({\rm PV})$ when the PV is reconstructed with and without the muon track.
The \xv and \mxip fits are performed independently at each mass, with the mean of the \ngob results used as the nominal value and half the difference assigned as a systematic uncertainty.

Both fit quantities are built from features that approximately follow \chisq probability density functions (PDFs) with minimal mass dependence.
The prompt-dimuon PDFs are taken directly from data at $m(\jpsi)$ and $m(Z)$, where prompt resonances are dominant (see Fig.~\ref{fig:m_prompt}).
Small \pt-dependent corrections are applied to obtain the PDFs at all other masses.
These PDFs are validated near threshold, at $m(\phi)$, and at $m(\Upsilon(1S))$, where the data predominantly consist of prompt dimuons.
The sum of the $hh$ and $h\mu_Q$ contributions, which each involve misidentified prompt hadrons, is determined using same-sign $\mu^{\pm}\mu^{\pm}$ candidates that satisfy all of the prompt-like criteria.
A correction is applied to the observed $\mu^{\pm}\mu^{\pm}$ yield at each mass to account for the difference in the production rates of $\pi^{+}\pi^{-}$ and $\pi^{\pm}\pi^{\pm}$, since double misidentified $\pi^{+}\pi^{-}$ pairs are the dominant source of the $hh$ background.
This correction, which is derived using a
prompt-like dipion data sample weighted by \pt-dependent muon-misidentification probabilities,
is as large as a factor of two near $m(\rho)$ but negligible for $\mmm \gtrsim 2\gev$.
The PDFs for the $\mu_Q\mu_Q$ background, which involves muon pairs produced in $Q$-hadron decays that occur displaced from the PV,
are obtained from simulation.
These muons are rarely produced at the same spatial point unless the decay chain involves charmonium.
Example \mxip fit results are provided in Ref.~\cite{Supp},  while Fig.~\ref{fig:m_prompt} shows the resulting data categorizations.
Finally, the \ngob yields are corrected for bin migration due to bremsstrahlung,
and the small expected Bethe-Heitler contribution is subtracted~\cite{Ilten:2016tkc}.

The prompt-like mass spectrum is scanned in steps of \sa/2 searching for \atomm contributions.
At each mass, a binned extended maximum likelihood fit is performed using all prompt-like candidates in a $\pm 12.5\sa$ window around \ma.
The profile likelihood is used to determine the $p$-value and the confidence interval for \naob, from which an upper limit at  90\% confidence level (CL) is obtained.
The signal PDFs are determined using a combination of simulated \atomm decays and the widths of the large resonance peaks observed in the data.
The strategy proposed in Ref.~\cite{BumpHunt} is used to select the background model and assign its uncertainty.
This method takes as input a large set of potential background components, which here includes all Legendre modes up to tenth order and dedicated terms for known resonances, and then performs a data-driven model-selection process whose uncertainty is included in the profile likelihood following Ref.~\cite{Dauncey:2014xga}.
More details about the fits, including discussion on peaking backgrounds, are provided in Ref.~\cite{Supp}.
The most significant  excess is $3.3\sigma$ at $\ma \approx 5.8\gev$, corresponding to a $p$-value of 38\% after accounting for the trials factor due to the number of prompt-like signal hypotheses.

\begin{figure}
  \centering
  \includegraphics[width=0.99\textwidth]{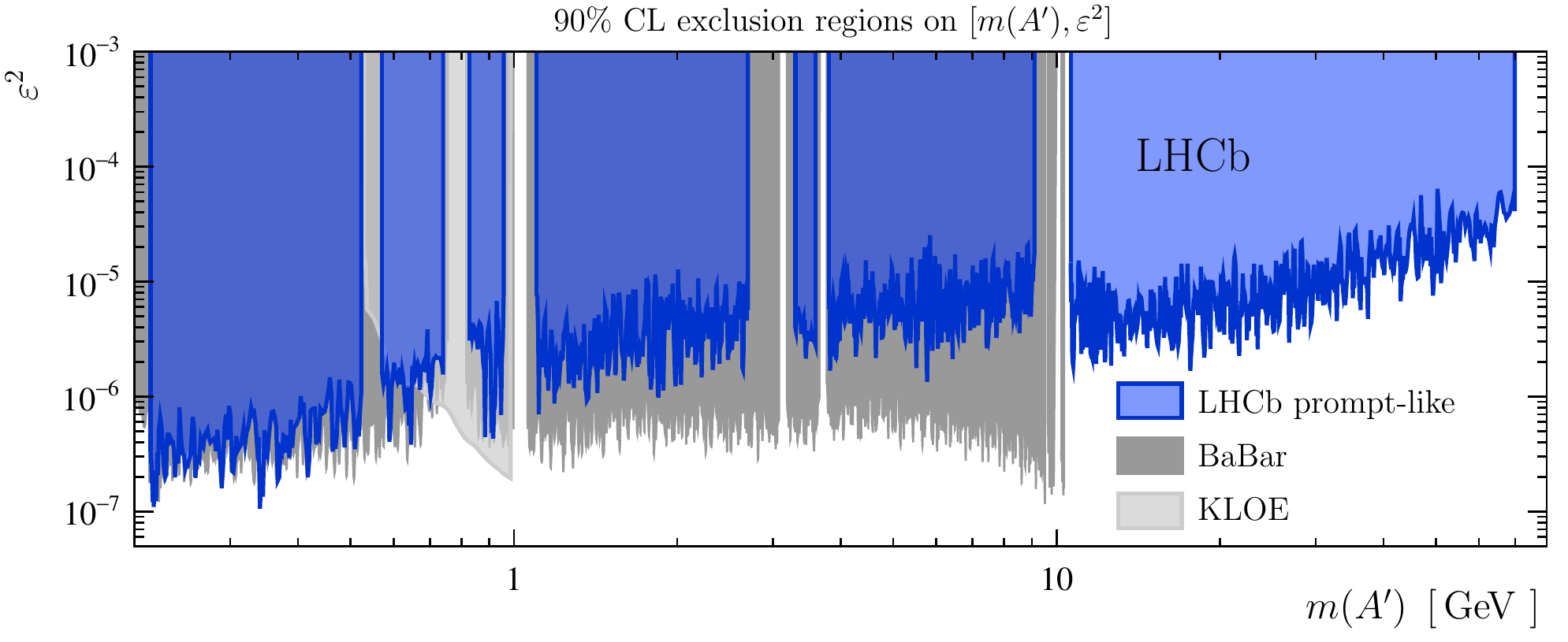}
  \caption{
  Regions of the \maeps parameter space excluded at 90\% CL by the  prompt-like \aprime search compared to the best existing limits~\cite{Lees:2014xha,KLOE:2016lwm}.
  }
  \label{fig:lims_prompt}
\end{figure}

Regions of the \maeps parameter space where the upper limit on \naob is less than \naex are excluded at 90\% CL.
Figure~\ref{fig:lims_prompt} shows that the constraints placed on prompt-like dark photons are comparable to the best existing limits below 0.5\gev, and are the most stringent for $10.6 < \ma < 70\gev$.
In the latter mass range, a nonnegligible model-dependent mixing with the $Z$ boson introduces additional kinetic-mixing parameters altering Eq.~\ref{eq:norm}; however, the expanded \aprime model space is highly constrained by precision electroweak measurements.
This search adopts the parameter values suggested in Refs.~\cite{Cassel:2009pu,Cline:2014dwa}.
The LHCb detector response is found to be independent of which quark-annihilation process produces the dark photon above 10\gev, making it easy to recast the results in Fig.~\ref{fig:lims_prompt} for other models.

For the long-lived dark photon search,
the stringent criteria applied in the trigger make contamination from prompt muon candidates negligible.
The dominant background contributions to the long-lived \aprime search are as follows:
photon conversions to $\mu^+\mu^-$ in the silicon-strip vertex detector (the VELO) that surrounds the $pp$ interaction region~\cite{LHCb-DP-2014-001};
$b$-hadron decays where two muons are produced in the decay chain;
and the low-mass tail from $\KS\to\pi^+\pi^-$ decays, where both pions are misidentified as muons.
Additional sources of background are negligible, {\em e.g.}\ kaon and hyperon decays, and $Q$-hadron decays producing a muon and a hadron that is misidentified as a muon.

Photon conversions in the VELO dominate the long-lived data sample at low masses.
A new method, which is described in detail in Ref.~\cite{VeloMat}, was recently developed for identifying particles created in secondary interactions with the VELO material.
A high-precision three-dimensional material map was produced from a data sample of secondary hadronic interactions.
Using this material map, along with properties of the \atomm decay vertex and muon tracks, a $p$-value is assigned to the photon-conversion hypothesis for each long-lived \atomm candidate.
A mass-dependent requirement is applied to these $p$-values that reduces the expected photon-conversion yields to a negligible level.

A characteristic signature of muons produced in $b$-hadron decays is the presence of additional displaced tracks.
Events are rejected if they are selected by the inclusive $Q$-hadron software trigger~\cite{Likhomanenko:2015aba} independently of the presence of the \atomm candidate.
Furthermore, two boosted decision tree (BDT) classifiers, originally developed for studying $B_{(s)}^0\to\mu^+\mu^-$ decays~\cite{LHCb-PAPER-2017-001},
are used to identify other tracks in the event that are consistent with having originated from the same $b$-hadron decay as the signal muon candidates.
The requirements placed on the BDT responses, which are optimized using a data sample of \KS decays as a signal proxy,
reject 70\% of the $b$-hadron background at a cost of about 10\% loss in signal efficiency.

As in the prompt-like \aprime search, the normalization is based on Eq.~\ref{eq:norm};
however, in the long-lived \aprime search, \effr is not unity, in part because the efficiency depends on the decay time, $t$.
Furthermore, the looser kinematic, muon-identification, and hardware-trigger requirements applied to long-lived \atomm candidates, {\em cf.}\ prompt-like candidates,  increase the efficiency by a factor of 7 to 10, ignoring $t$-dependent effects.
These \ma-dependent factors are determined using a small control data sample of dimuon candidates consistent with originating from the PV, but otherwise satisfying the long-lived criteria.
A relative 10\% systematic uncertainty is assigned to the long-lived \atomm normalization due to background contamination in the control sample.

The fact that the kinematics are identical for \atomm and prompt \gtomm decays for $\ma = m(\gamma^*)$ enables the $t$ dependence of the signal efficiency to be determined using a data-driven approach.
For each value of $[\ma,\ta]$, prompt \gtomm candidates in the control data sample near \ma are resampled many times as long-lived \atomm decays, and all $t$-dependent properties, {\em e.g.}\ \mxip,
are recalculated based on the resampled decay-vertex locations.
This approach is validated in simulation by using prompt \atomm decays to predict the properties of long-lived \atomm decays, and based on these studies a 2\% systematic uncertainty is assigned to the signal efficiencies.
The $\effr$ values integrated over $t$ are provided in Ref.~\cite{Supp}.

A scan is again performed in discrete steps of \sa/2 looking for \atomm contributions; however, in this case, discrete steps in \ta are also considered.
Binned extended maximum likelihood fits are performed using all long-lived candidates and the three-dimensional feature space of \mmm, $t$,
and the consistency of the decay topology as quantified in the decay-fit $\chi^2_{\rm DF}$, which has three degrees of freedom (the data distribution is provided in Ref.~\cite{Supp}).
The expected conversion contribution is derived in each bin from the number of candidates rejected by the conversion criterion.
Two large control data samples are used to develop and validate the modeling of the $b$-hadron and \KS contributions:
candidates that fail the $b$-hadron suppression requirements,
and
candidates that fail but nearly satisfy the muon-identification requirements.
The profile likelihood is used to obtain the $p$-values and confidence intervals on \naobt.
The most significant excess occurs at $\ma = 239\mev$ and $\ta = 0.86\ps$, where the $p$-value corresponds to $3.0\sigma$.
Considering only the long-lived-search trials factor reduces this to $2.0\sigma$.
More details about these fits are provided in Ref.~\cite{Supp}.

\begin{figure}
  \centering
  \includegraphics[width=0.69\textwidth]{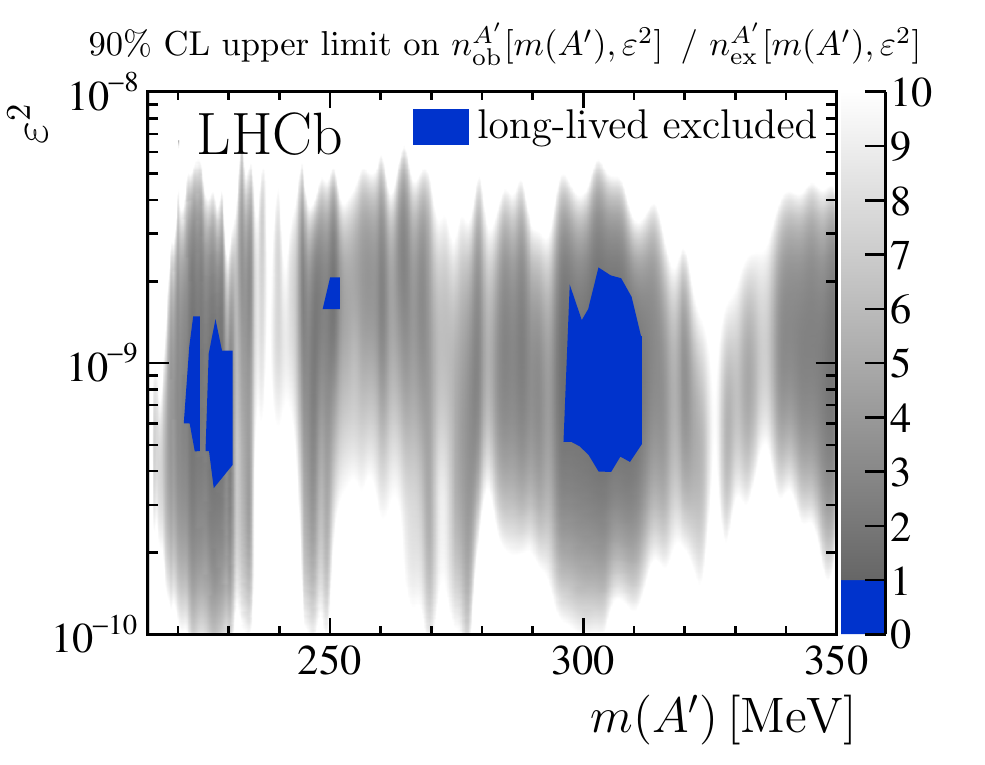}
  \caption{
  Ratio of the observed upper limit on \naobe at 90\% CL to its expected value, where regions less than unity are excluded. There are no constraints from previous experiments in this region.
  }
  \label{fig:displ_lims}
\end{figure}

Under the assumption that \aprime decays to invisible final states are negligible, there is a fixed (and known) relationship between \ta and $\varepsilon^2$ at each mass~\cite{Ilten:2016tkc}; therefore, the upper limits on \naobt can be translated into limits on \naobe.
Regions of the \maeps parameter space where the upper limit on \naobe is less than \naex are excluded at 90\% CL (see Fig.~\ref{fig:displ_lims}).
While only small regions of $[\ma,\varepsilon^2]$ space are excluded, a sizable portion of this parameter space will soon become accessible as more data are collected.

In summary, searches are performed for both prompt-like and long-lived dark photons produced in $pp$ collisions at a center-of-mass energy of 13\tev,
using \atomm decays and a data sample corresponding to an integrated luminosity of 1.6\invfb collected with the LHCb detector during 2016.
The prompt-like \aprime search covers the mass range from near the dimuon threshold up to 70\gev, while the long-lived \aprime search is restricted to the low-mass region $214<\ma<350\mev$.
No evidence for a signal is found, and 90\% CL exclusion regions are set on the $\gamma$--$\aprime$ kinetic-mixing strength.
The constraints placed on prompt-like dark photons are the most stringent to date for the mass range $10.6 < \ma < 70\gev$, and are comparable to the best existing limits for $\ma < 0.5\gev$.
The search for long-lived dark photons is the first to achieve sensitivity using a displaced-vertex signature.

These results demonstrate the unique sensitivity of the LHCb experiment to dark photons, even using a data sample collected with a trigger that is inefficient for low-mass \atomm decays.
Using knowledge gained from this analysis, the software-trigger efficiency for low-mass dark photons has been significantly improved for 2017 data taking.
Looking forward to Run~3, the planned increase in luminosity and removal of the hardware-trigger stage should increase the number of expected \atomm decays in the low-mass region by a factor of $\mathcal{O}(100$--1000$)$ compared to the 2016 data sample.

\section*{Acknowledgements}

\noindent We express our gratitude to our colleagues in the CERN
accelerator departments for the excellent performance of the LHC. We
thank the technical and administrative staff at the LHCb
institutes. We acknowledge support from CERN and from the national
agencies: CAPES, CNPq, FAPERJ and FINEP (Brazil); MOST and NSFC
(China); CNRS/IN2P3 (France); BMBF, DFG and MPG (Germany); INFN
(Italy); NWO (The Netherlands); MNiSW and NCN (Poland); MEN/IFA
(Romania); MinES and FASO (Russia); MinECo (Spain); SNSF and SER
(Switzerland); NASU (Ukraine); STFC (United Kingdom); NSF (USA).  We
acknowledge the computing resources that are provided by CERN, IN2P3
(France), KIT and DESY (Germany), INFN (Italy), SURF (The
Netherlands), PIC (Spain), GridPP (United Kingdom), RRCKI and Yandex
LLC (Russia), CSCS (Switzerland), IFIN-HH (Romania), CBPF (Brazil),
PL-GRID (Poland) and OSC (USA). We are indebted to the communities
behind the multiple open-source software packages on which we depend.
Individual groups or members have received support from AvH Foundation
(Germany), EPLANET, Marie Sk\l{}odowska-Curie Actions and ERC
(European Union), ANR, Labex P2IO, ENIGMASS and OCEVU, and R\'{e}gion
Auvergne-Rh\^{o}ne-Alpes (France), RFBR and Yandex LLC (Russia), GVA,
XuntaGal and GENCAT (Spain), Herchel Smith Fund, the Royal Society,
the English-Speaking Union and the Leverhulme Trust (United Kingdom).

\setboolean{inbibliography}{true}
\bibliographystyle{LHCb}
\bibliography{main,LHCb-PAPER,LHCb-CONF,LHCb-DP,LHCb-TDR,PRL}

\newpage

\section*{Supplemental Material}

\subsection*{Prompt-Like Fits}

The fit strategy denoted by aic-o and described in detail in Ref.~\cite{BumpHunt} is used in the prompt-like \aprime search.
The \mmm spectrum is scanned in steps of \sa/2 searching for \atomm contributions.
At each mass, a binned extended maximum likelihood fit is performed, and the profile likelihood is used to determine the $p$-value and the confidence interval on \naob.
The prompt-like-search trials factor is obtained using pseudoexperiments.
As in Ref.~\cite{BumpHunt}, each fit is performed in a $\pm12.5\sa$ window around the scan-mass value using bins with widths of $\sa/20$.
Near threshold, the quantity $q(\mu^+\mu^-)\equiv \sqrt{\mmm^2 - 4m(\mu)^2}$ is used instead of the mass since it is easier to model.
The confidence intervals are defined using the {\em bounded likelihood} approach,
 which involves taking $\Delta \log{\mathcal{L}}$ relative to zero signal, rather than the best-fit value, if the best-fit signal value is negative.
This approach enforces that only physical (nonnegative) upper limits are placed on \naob, and prevents defining exclusion regions that are much better than the experimental sensitivity in cases where a large deficit in the background yield is observed.

The signal models are determined at each \ma using a combination of simulated \atomm decays and the widths of the large resonance peaks that are clearly visible in the data.
The background models are chosen following the method of Ref.~\cite{BumpHunt}.
This method takes as input a large set of potential background components, then performs a data-driven model-selection process whose uncertainty is included in the profile likelihood following Ref.~\cite{Dauncey:2014xga}.
In this analysis, the set of possible background components includes all Legendre modes with $\ell \leq 10$ at every \ma.
Additionally, dedicated background components are included to model the near-threshold turn-on behavior and all sizable known resonance contributions.

The use of 11 Legendre modes adequately describes every double-misidentified peaking background that contributes at a significant level, {\em e.g.}, $\phi\to K^+K^-$ and ${D\to K^{\pm}\pi^{\mp}}$ double misidentified as dimuons, and in the $D$ case misreconstructed as prompt-like, do not require dedicated background components.
In mass regions where such complexity is not required, the data-driven model-selection procedure reduces the complexity which increases the sensitivity to a potential signal contribution.
As in Ref.~\cite{BumpHunt}, all fit regions are transformed onto the interval $[-1,1]$, where the scan \ma value maps to zero.
After such a transformation, the signal model is (approximately) an even function;
therefore, odd Legendre modes are orthogonal to the signal component, which means that the presence of odd modes has minimal impact on the variance of \naob.
In the prompt-like fits, all odd Legendre modes up to ninth order are included in every background model, while only a subset of the even modes is selected for inclusion in each fit.

Regions in the mass spectrum where large known resonance contributions are observed are vetoed in the prompt-like \aprime search.
Furthermore, the regions near the $\eta^{\prime}$ meson and the excited $\Upsilon$ states (beyond the $\Upsilon(4S)$ meson) are treated specially.
For example, since it is not possible to distinguish between \atomm and $\eta^{\prime}\!\to\!\mu^+\mu^-$ contributions at $m(\eta^{\prime})$, the $p$-values near this mass are ignored.
Any excess at $m(\eta^{\prime})$ is treated as signal when setting the limits on \naob, which is conservative in that a $\eta^{\prime}\!\to\!\mu^+\mu^-$ contribution will weaken the constraints on \atomm decays.
The same strategy is used near the excited $\Upsilon$ masses.
The treatment of all mass regions is summarized in Table~\ref{tab:prompt_fit}.

\begin{table}[h!]
  \begin{center}
    \caption{\label{tab:prompt_fit}
    Summary of mass regions with special treatment in the prompt-like \aprime search. In all other mass regions from 214\mev to 70\gev, limits are set and the $p$-values are considered as possible evidence for \atomm decays.}
      \begin{tabular}{ccc}
        \toprule
    \ma Region [\mev] & Resonance(s) & Special Treatment \\
        \midrule
        $\phantom{00}(524,571)\phantom{00}$ & $\eta$  & no search  \\
        $\phantom{00}(741,827)\phantom{00}$ & $\omega$ & no search  \\
        $\phantom{00}(940,960)\phantom{00}$ & $\eta^{\prime}$  & limits set, but $p$-values ignored \\
        $\phantom{00}(960,1100)\phantom{0}$ & $\phi$ & no search  \\
        $\phantom{0}(2700,3300)\phantom{0}$ & \jpsi  & no search  \\
        $\phantom{0}(3600,3800)\phantom{0}$ & $\psi(2S)$ and $\psi(3770)$ & no search  \\
        $\phantom{0}(9100,10600)$ & $\Upsilon(1S)$--$\Upsilon(4S)$ & no search  \\
        $(10840,11040)$ & excited $\Upsilon$ states & limits set, but $p$-values ignored \\
        \bottomrule
      \end{tabular}
  \end{center}
\end{table}

\subsection*{Long-Lived Fits}

\begin{figure}
  \centering
  \includegraphics[width=0.7\textwidth]{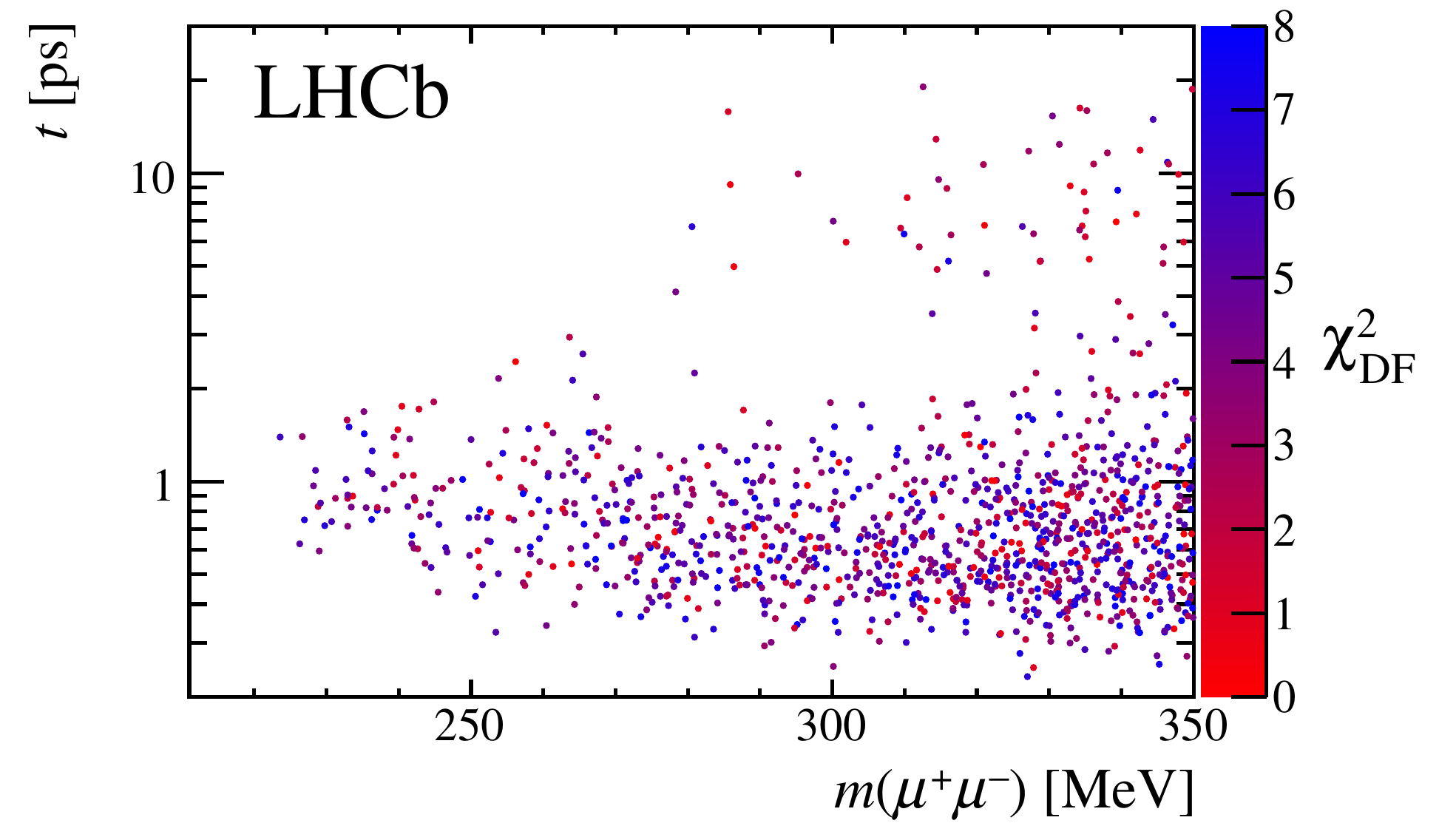}
  \caption{
  Three-dimensional distribution of $\chi^2_{\rm DF}$ versus $t$ versus \mmm, which is fit to determine the long-lived signal yields.
  The data are consistent with being predominantly due to $b$-hadron decays at small $t$, and due to \KS decays for large $t$ and $\mmm \gtrsim 280\mev$.
  The largest signal-like excess occurs at $\ma = 239\mev$ and $\ta = 0.86\ps$.
  }
  \label{fig:displ}
\end{figure}

The long-lived signal yields are determined from binned extended maximum likelihood fits performed on all long-lived \atomm candidates using the three-dimensional feature space of the dimuon invariant mass, \mmm, the \aprime decay time, $t$, and the decay-fit quality, $\chi^2_{\rm DF}$.
As in the prompt-like \aprime search, a scan is performed in discrete steps of \sa/2; however, in this case, discrete steps in \ta are also considered.
The profile likelihood is again used to obtain the $p$-values and the confidence intervals on \naobt.
The binning scheme involves four bins in $\chi^2_{\rm DF}$: [0,2], [2,4], [4,6], and [6,8].
Eight bins in $t$ are used: [0.2,0.6], [0.6,1.1], [1.1,1.6], [1.6,2.2], [2.2,3.0], [3,5], [5,10], and $>\!10\ps$.
The binning scheme used for \mmm depends on the scan \ma value, and is chosen such that the majority of the signal falls into a single bin.
Signal decays mostly have small $\chi^2_{\rm DF}$ values, with about 50\%\,(80\%) of \atomm decays satisfying $\chi^2_{\rm DF} < 2\,(4)$.
Background from $b$-hadron decays populates the small $t$ region and is roughly uniformly distributed in $\chi^2_{\rm DF}$, whereas background from \KS decays is signal-like in $\chi^2_{\rm DF}$ and roughly uniformly distributed in $t$.
Figure~\ref{fig:displ} shows the three-dimensional distribution of all long-lived \atomm candidates.

The expected contribution in each bin from photon conversions is derived from the number of candidates rejected by the conversion criterion.
As discussed in the Letter, two large control data samples are used to develop and validate the modeling of the $b$-hadron and \KS contributions.
Both contributions are well modeled by the function $\Theta[q(\mu^+\mu^-)-q_0]\times\{a[q(\mu^+\mu^-)-q_0]+b[q(\mu^+\mu^-)-q_0]^2\}$, where $q_0$, $a$, and $b$ are fitted to the data, and $\Theta$ denotes the Heaviside step function.
While no evidence for $t$ or $\chi^2_{\rm DF}$ dependence is observed for these parameters in either the $b$-hadron or \KS control sample, all parameters are allowed to vary independently in each $[t,\chi^2_{\rm DF}]$ region in the fits used in the long-lived \aprime search.

\begin{figure}[h!]
  \centering
  \includegraphics[width=0.49\textwidth]{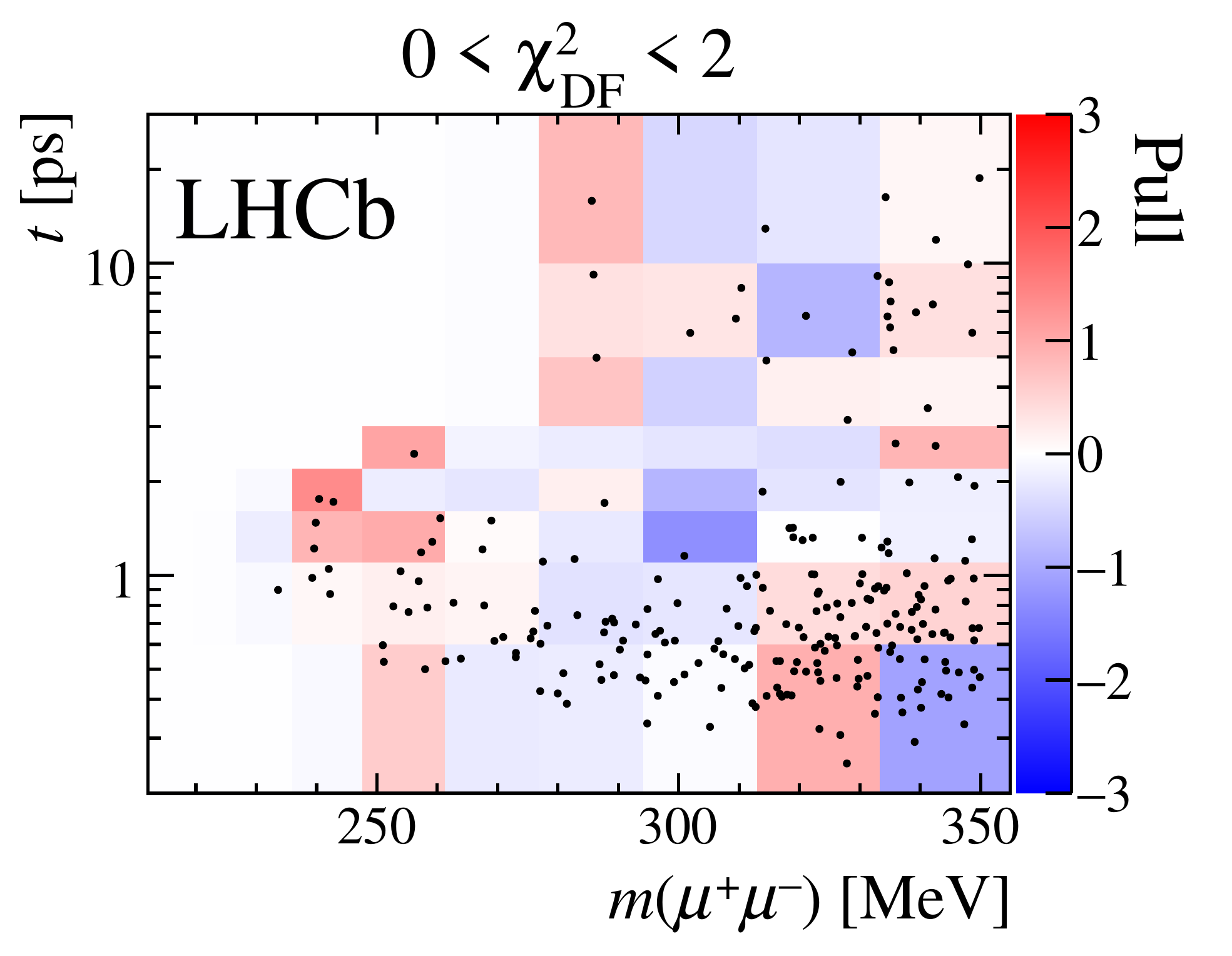}
  \includegraphics[width=0.49\textwidth]{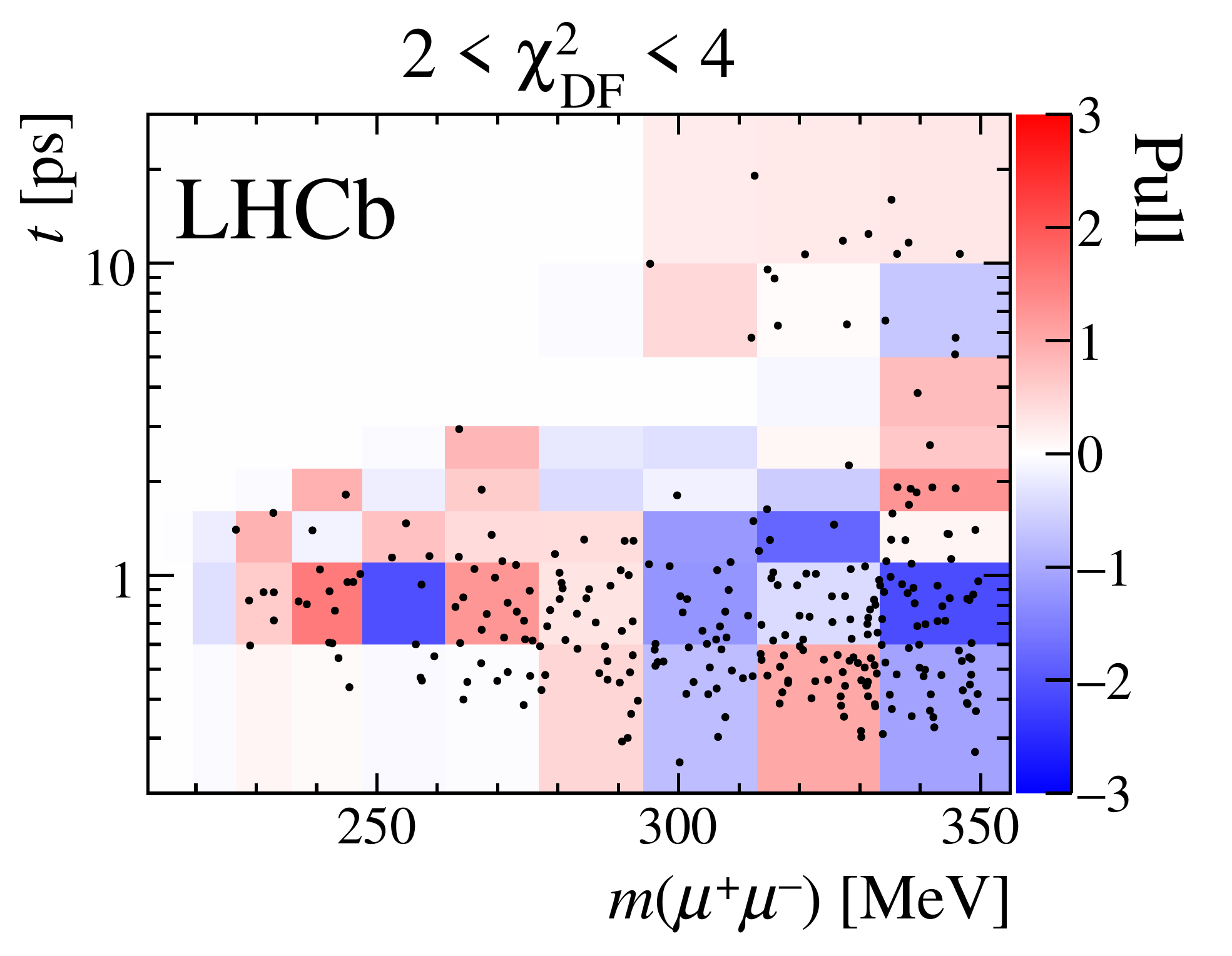}
  \includegraphics[width=0.49\textwidth]{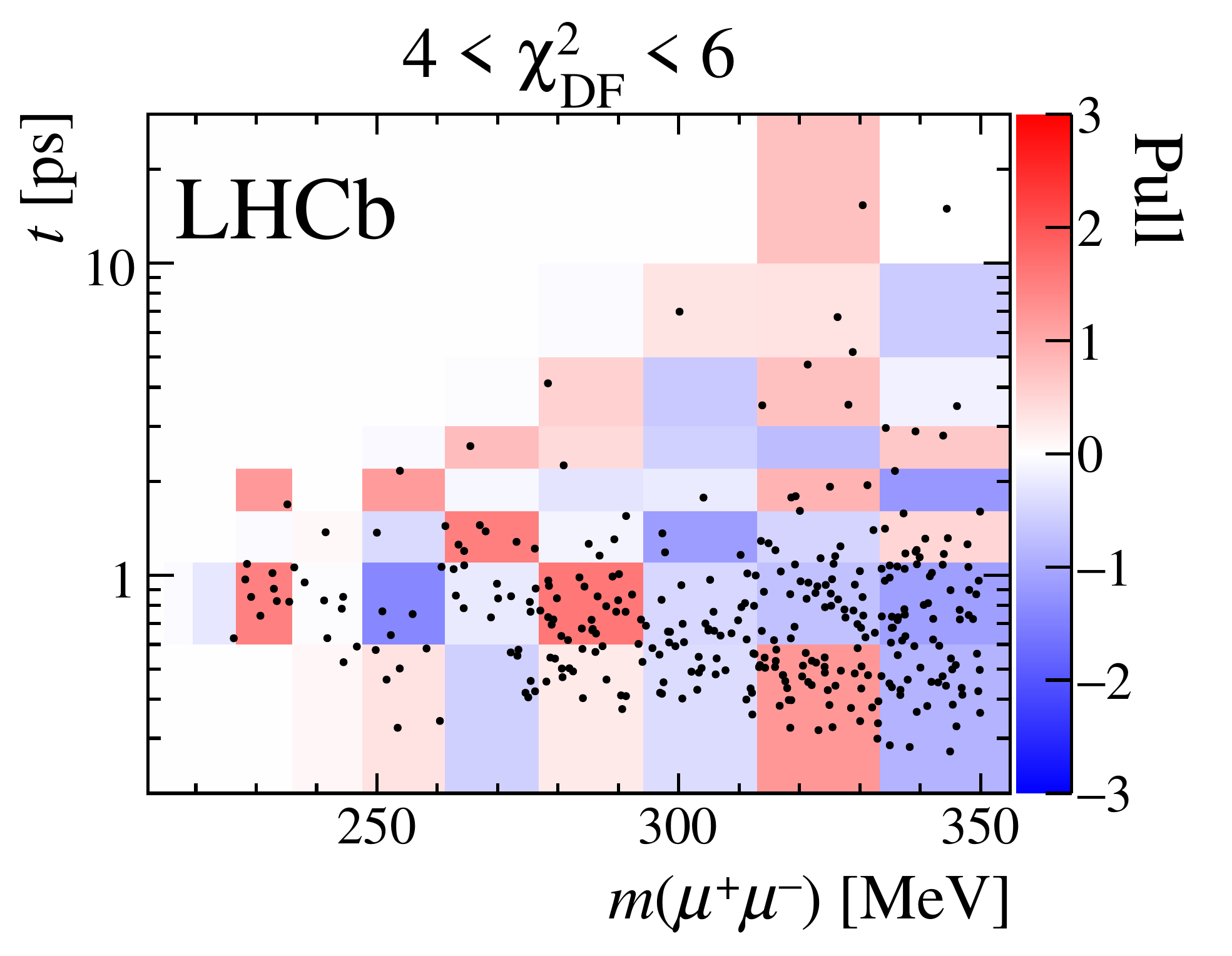}
  \includegraphics[width=0.49\textwidth]{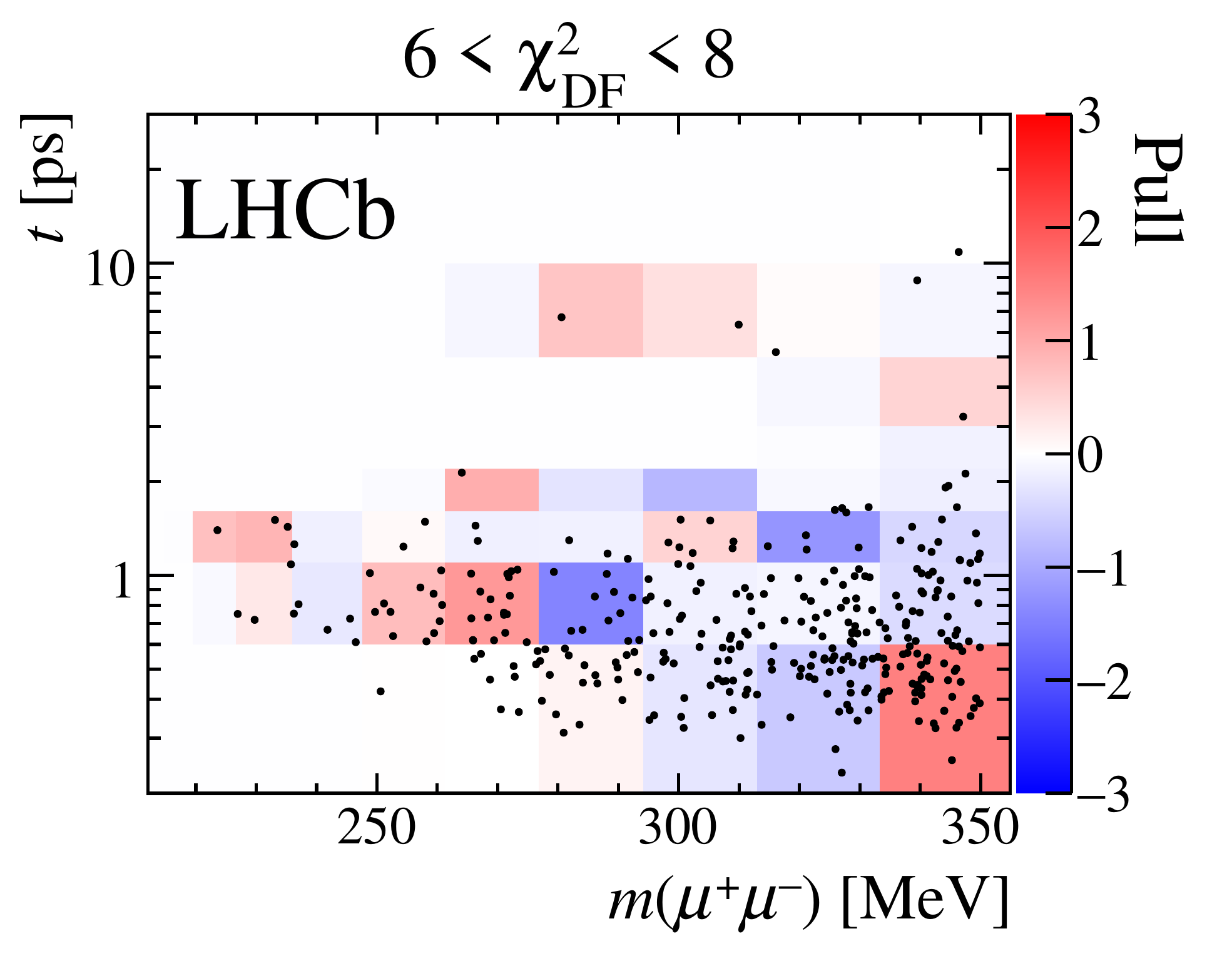}
  \caption{
  Long-lived \atomm candidates (black points) showing $t$ versus \mmm in bins of $\chi^2_{\rm DF}$, compared to the pulls from binned fits performed without a signal contribution (color axis).
  Positive pulls denote an excess of data candidates.
  }
  \label{fig:displ_pulls}
\end{figure}

Figure~\ref{fig:displ_pulls} shows the long-lived \atomm candidates, along with the pull values obtained from fits performed to the data where no signal contributions are included.
All of the pulls are in the range $[-2,2]$.
{\em N.b.}, due to the fact that the background threshold parameters are free to vary in each $[t,\chi^2_{\rm DF}]$ region, the lowest-mass nonempty bin for each $[t,\chi^2_{\rm DF}]$ is biased towards a positive pull in the absence of a signal contribution.

\begin{figure}[h!]
  \centering
  \includegraphics[width=0.7\textwidth]{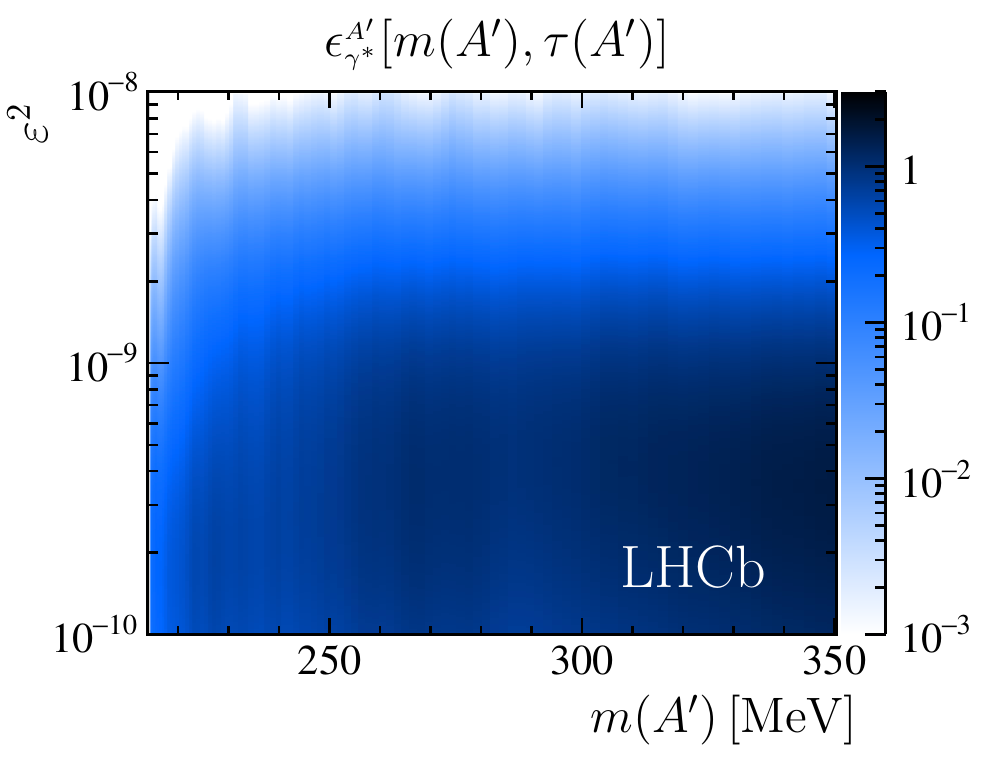}
  \caption{
  Efficiency ratio \effr for long-lived dark photons, integrated over decay time. The sharp decrease at larger values of $\varepsilon^2$ is due to the stringent \mxip criterion applied in the 2016 trigger.
  }
  \label{fig:rel_eff}
\end{figure}

\clearpage

\subsection*{Additional Figures}

\begin{figure}[h!]
  \centering
  \includegraphics[width=0.99\textwidth]{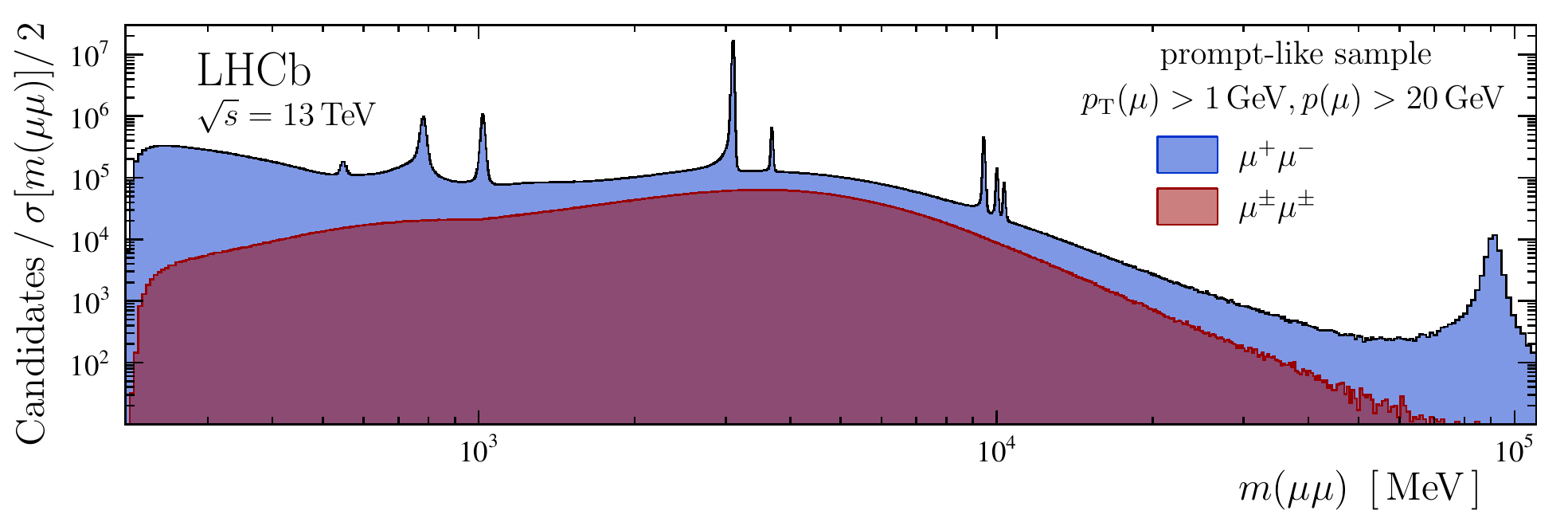}
  \caption{
  Mass spectrum selected by the prompt-like \atomm trigger.
  }
  \label{fig:trig}
\end{figure}

\begin{figure}[h!]
  \centering
  \includegraphics[width=0.32\textwidth]{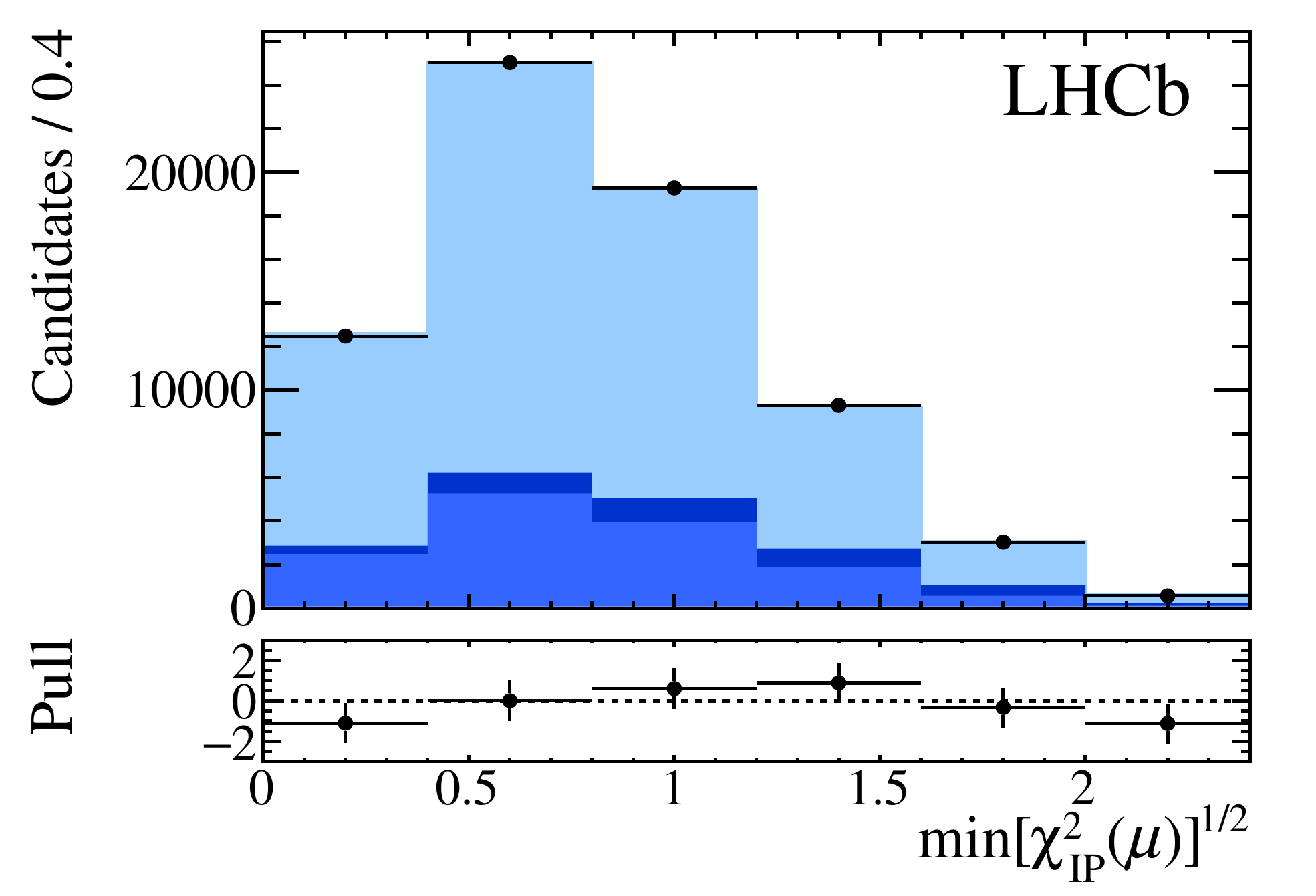}
  \includegraphics[width=0.32\textwidth]{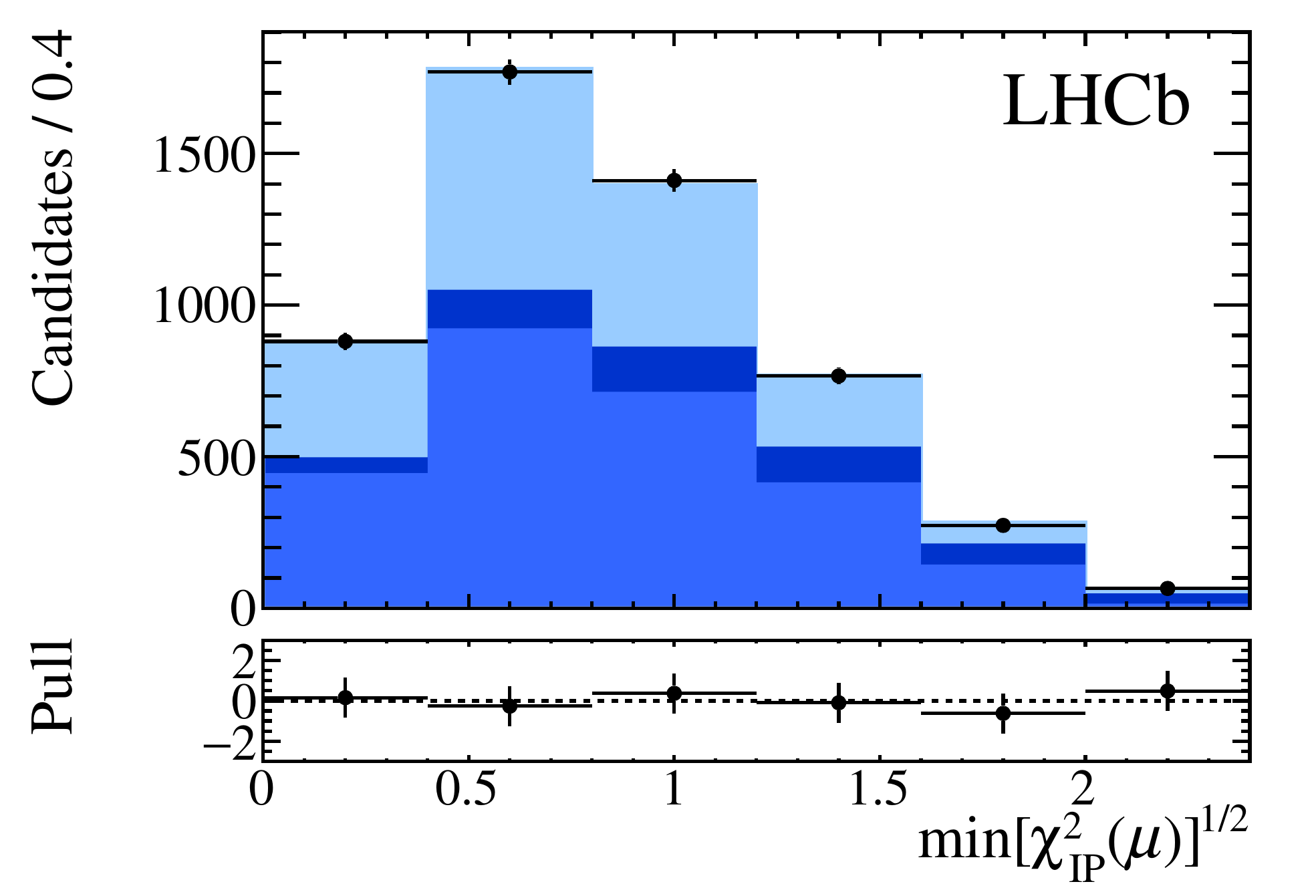}
  \includegraphics[width=0.32\textwidth]{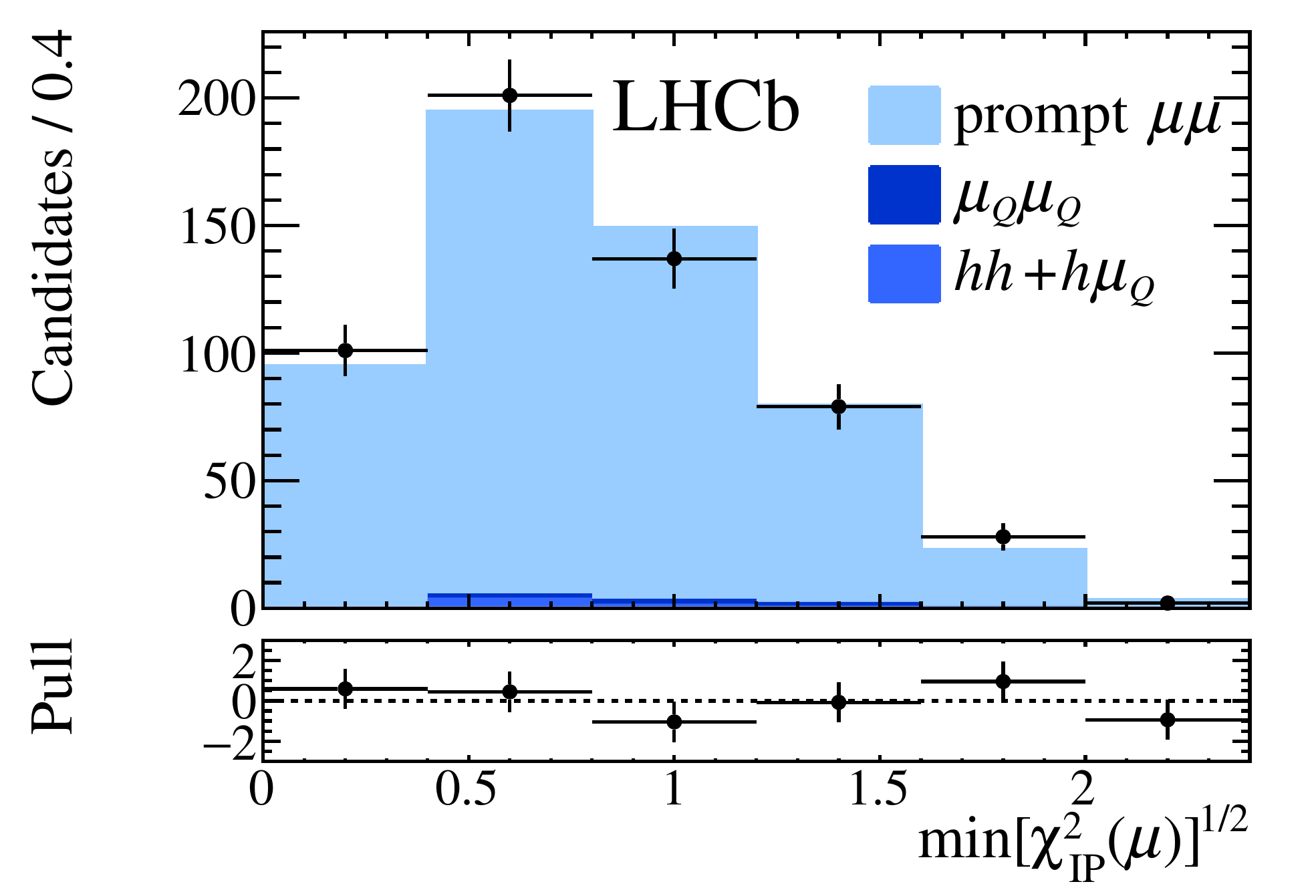}
  \caption{
  Example $\mxip^{1/2}$ distributions with fit results overlaid for prompt-like candidates near (left) $\ma=0.5$, (middle) 5, and (right) 50\gev. The square root of \mxip is used in the fits to increase the bin occupancies at large \mxip values.
  }
  \label{fig:fits_prompt}
\end{figure}

\begin{figure}[h!]
  \centering
  \includegraphics[width=0.99\textwidth]{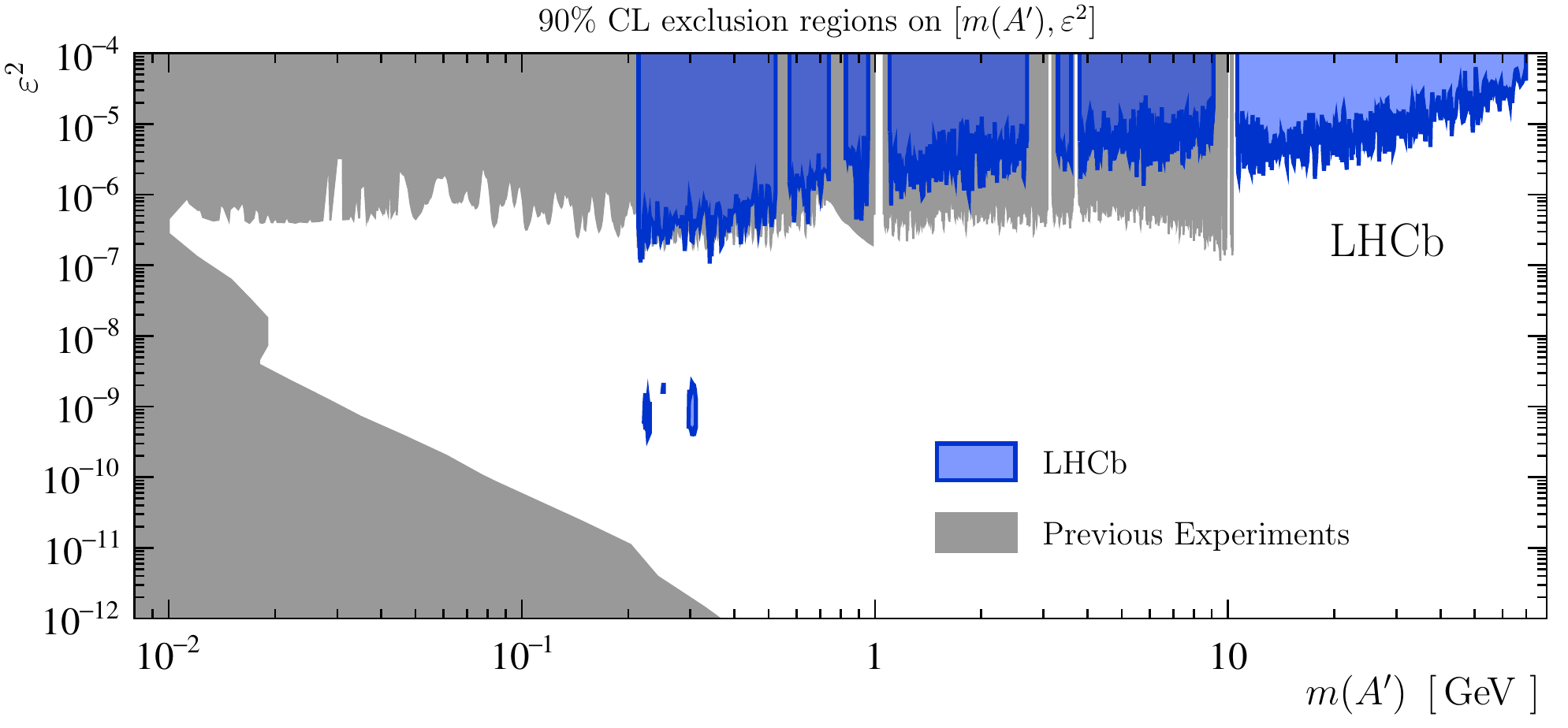}
  \caption{
  Comparison of the results presented in this Letter to existing constraints from previous experiments (see Ref.~\cite{Alexander:2016aln} for details about previous experiments). }
  \label{fig:xip_prompt}
\end{figure}

\clearpage

\centerline{\large\bf LHCb collaboration}
\begin{flushleft}
\small
R.~Aaij$^{40}$,
B.~Adeva$^{39}$,
M.~Adinolfi$^{48}$,
Z.~Ajaltouni$^{5}$,
S.~Akar$^{59}$,
J.~Albrecht$^{10}$,
F.~Alessio$^{40}$,
M.~Alexander$^{53}$,
A.~Alfonso~Albero$^{38}$,
S.~Ali$^{43}$
% ...

{\footnotesize \it
% ...
$ ^{5}$Clermont Universit{\'e}, Universit{\'e} Blaise Pascal, CNRS/IN2P3, LPC, Clermont-Ferrand, France\\
% ...
$ ^{10}$Fakult{\"a}t Physik, Technische Universit{\"a}t Dortmund, Dortmund, Germany\\
% ...
$ ^{38}$ICCUB, Universitat de Barcelona, Barcelona, Spain\\
$ ^{39}$Universidad de Santiago de Compostela, Santiago de Compostela, Spain\\
$ ^{40}$European Organization for Nuclear Research (CERN), Geneva, Switzerland\\
% ...
$ ^{43}$Nikhef National Institute for Subatomic Physics, Amsterdam, The Netherlands\\
% ...
$ ^{48}$H.H. Wills Physics Laboratory, University of Bristol, Bristol, United Kingdom\\
% ...
$ ^{53}$School of Physics and Astronomy, University of Glasgow, Glasgow, United Kingdom\\
% ...
$ ^{59}$University of Cincinnati, Cincinnati, OH, United States\\
% ...
}
\end{flushleft}

\end{document}